# Multiresolution Division Multiplex (MRDM): A New Wavelet-based Multiplex System

H. M. de Oliveira, Eric A. Bouton

*Abstract*—An original multiplex scheme is introduced, which is based on Mallat's multiresolution formulation of wavelet systems. This system is adaptable and its implementation is well matched to digital signal processors and computers. The approach termed multiresolution division multiplex (MRDM) is intensive in signal processing (SP) tools, extremely flexible and can combine a variety of tributaries at different bit rates. A broad variety of orthogonal wavelet systems $\{\varphi_k, \psi_{k,j}\}$ can endow with MRDM and the channel waveforms, and consequently the spectral shape and system performance depend upon the selected wavelets. Demultiplex can be done efficiently, since the number of floating multiplications and additions increase only linearly with the length of signals. A Haar-based MRDM scheme is presented to illustrate the versatility of this new multiplex approach.

## I. PRELIMINARIES AND HISTORY

MULTIPLEX and multiple access techniques are nowadays central piece of nearly all modern communication systems. The term multiplex came from the Multiplexor apparatus, conceived around 1876 by French telegraph engineer Jean Émile Baudot, which could achieve the simultaneous transmission of six telegraph signals over a single telegraph channel. This technique was basically kept only for telegraphy until the beginning of twenty century, when the amazing telephony growth required introducing multiplex for voice signals. The emergence of modulation techniques, Single Side Band (SSB) in particular, produced a new multiplex system: The Frequency Division Multiplex (FDM), which rapidly reached a huge triumph in telephone exchange offices [1]. Historically, telephone networks used FDM to carry several voice channels on a single physical circuit since 1910. FDM systems quickly expanded through the world. In 1920, the telecommunications engineer Georges Valensi (ITT, Paris) launched the fundamentals of the Time Division Multiplex (TDM) for voice channels, although he is mostly recognized by his method of transmitting colour images so that they could be received on both colour and black & white television sets. According to Bell Labs history [1], he offered then his newborn idea to the America as a gift expressing his gratitude for aid in World War I (*sic*). The manuscript was afterwards analysed by J. Carson, who correctly argued that such an idea was infeasible for the state-of-the-art technology. The crux of matter of Valensi's idea was over again evoked by Maurice Deloraine (ITT, Paris) and John Bennett (Bell Labs), who independently rediscovery TDM. Both are here and there recognized as the inventors of the digital multiplexing [2], [3]. Ever since the colossal work of Claude E. Shannon [4], digital has been replacing standard techniques in communication systems and more. Digital multiplex concerns as a rule the Time Division Multiplex (TDM), both on the standard plesiochronous (PDH) and more recently on the synchronous transmission mode (SDH) [5]. However, digital multiplex can also be achieved by Coding Division Multiplex (CDM), which has recently been the focus of interest, especially after the IS-95 standardisation of the CDMA system for cellular telephone (superseded by the IS-2000 standard) [6] and W-CDMA [7]. Classical multiplex increases simultaneously the transmission rate and the bandwidth by the same factor, thereby keeping unchanged the spectral efficiency. In order to achieve (slight) better spectral efficiencies, classical CDMA uses waveforms presenting a nonzero but residual correlation. With the advent of optical fibres in communication systems, "another" multiplex technique was bringing to play. In fact, frequency division multiplexing in the optical domain is known as wavelength division multiplexing (WDM). In fibre optic telecommunications, WDM is a technology that multiplexes multiple optical carrier signals on a single optical fibre by using diverse wavelengths (colours) of laser light to carry different signals. In 1999, a new approach termed Galois-Division multiplex (GDM) was introduced as an efficient-bandwidth multiplex scheme for bandlimited channels [8]. It can be implemented along with fast transform algorithms and offer compact bandwidth requirements. GDM systems are truly digital multiplex schemes, which employ spread spectrum techniques based on finite field (FF) transforms [9], such as the FF Fourier transform introduced by Pollard [10]. Information strings coming from users are combined by means of alphabet expansion (resembling the prize winning coded-modulation technique [11]). The users' sequences *v* are defined over a finite field GF(*p*). Thus, the multiplex (mux) involves an expanded signal set whose symbols are defined over the extension field, GF($p^m$). Orthogonal Galois-field spreading sequences [12] are a new tool to perform multilevel direct sequence spread spectrum communication (DS-SS). Systems that employ Galois-field spreading sequences are the so-called Galois-Division Multiple Access (GDMA) [13]. Since signal-processing (SP) techniques have been experiencing fantastic advances [14], SP-based multiplex that extensively apply fast algorithms are welcome, particularly from the implementation point of view.

This work was partially supported by the Brazilian National Council for Scientific and Technological Development (CNPq) under research grant #306180 (HMdO). EAB is grateful to CNPq for an MSc scholarship. The authors are with the Signal Processing Group, Federal University of Pernambuco, C.P. 7800, CEP: 50711-970, Recife-PE, Brazil
 (e-mail: hmo@ufpe.br;ebouton@terra.com.br)

The more astonishing advances in modern SP are perhaps related to wavelet theory [15]. Specifically, wavelet emerged as influential SP tools for efficiently handling signals, especially due to its full computational power. If the input requires $N$ data point, the computational complexity of the wavelet transform operations are typically $O(N)$. Stéphane Mallat launched, in 1987, the multiresolution technique for signal decomposition [16]. Nearly all the organizational structures, including those of Biological sensory systems, are organized into "levels" or "scales". Therefore, the multiscale representation is an essential ingredient for efficiently extracting information from observation [17], [18]. The main aim of this paper is to set up an original multiplex technique, which held such features. These new schemes are referred hereafter to as Multiresolution Division Multiplex (MRDM) and advanced DSP families can be of assistance for implementing it. It is worth to state that our purpose here is barely to launch the key facts of these novel systems.

II. THE MULTIRESOLUTION MULTIPLEX: KEY CONCEPTS

According to Mallat's heterogeneous wavelet decomposition [16], given a wavelet system $\{\varphi_k, \psi_{k,j}\}$, a signal $f(t)$ can be written as

$$f(t) \cong \sum_k c_k \varphi_{Jk}(t) + \sum_k \sum_{j=1}^J d_{j,k} \psi_{jk}(t), \quad (1)$$

where $c_k$ and $d_{jk}$ are the approximation and detail coefficients of the decomposition. In this case, MRA is used as a decomposition of an intricate signal (analysis). Instead of using Mallat's decomposition to analyse a *single* given signal, this approach can directly generate the multiplexed signal from a number of users. The idea is to apply MRA so as to synthesize the multiplexed signal (synthesis). Let $f_i(t)$ be the analog signal from the $i^{th}$ user. If samples of each signal are attributed to a given scale, for instance,

$$c_k = f_0[k] \text{ and } d_{jk} = f_j[k], \quad (2)$$

a muxed continuous signal can be generated by building

$$\varphi_{MUX}(t) \cong \sum_k f_0[k]\varphi_{Jk}(t) + \sum_k \sum_{j=1}^J f_j[k]\psi_{jk}(t). \quad (3)$$

In this expression, in contrast with standard MRA, each coefficient came from a different user. This pooled-signal can easily be demultiplexed at the receiver by Mallat's algorithm that recover the scale and wavelet coefficients of $\varphi_{MUX}(t)$, which are in fact users' signal samples. A "pseudo-MRA" synthesis is used to engender the multiplexed signal and demultiplex is performed through an MRA decomposition of the multiplexed signal. This kind of multiplex technique is referred so as to MRDM.

*A. Bandlimited analog signal multiplexed*

In accordance with the multiresolution theory, the lengths of the approximation and details at different scales are not the same. Bearing in mind a multiresolution system with $J$ levels of decomposition, the rates of the $(J+1)$-users are not identical and users' rate is scale-dependent. A given user accesses just a single level of the pseudo-MRA ($J$ channels set detail coefficients and 1 channel set scale coefficients).

Considering a first case (that is, analog users at different sample rate), let $N$ be the blocklength of the discrete time signal at the multiplex output. Let $J \leq \log_2 N$ be the number of scales of the pseudo-MRA. The different signal lengths are $N/2^j$; $j = 1,2...J$. The analog input MRDM scheme is based on the Shannon-Nyquist-Koteln'kov sampling theorem (SNK theorem) so it is applied to bandlimited signals. If $J$ is the number of scales in the pseudo-MRA, the mux can handle with signals of maximum frequency $f_m, 2f_m, 4f_m, \ldots, 2^{J-1}f_m$. The value of the wavelet coefficients $f_j[k]$ is set according to a sample of signal of the $j^{th}$-channel.

*B. Digital signals multiplexed*

In a digital input MRDM, a D/A converter is used and the bit stream of each channel is segmented according to the number of bits of the converter. Each binary word yields a real-valued sample that is assigned to the (scale or wavelet) coefficient $f_j[k]$. The whole MRDM process can be better understood through a small number of naïve examples shown in the sequel.

Considering an initial case, where all users have identical rate, the understanding of the system philosophy becomes easier. A few parameters are preset to put the method to work: the number of levels in the pseudo-MRA synthesis ($J$) and the length of the information blocks from each user ($N/2^J$). The total number of $n_J=2^J$ simultaneous tributaries of equal rate is distributed through the pseudo-MRA according to the following design:

| | |
|---|---|
| 1st level | $2^{J-1}$ tributaries |
| 2nd level | $2^{J-2}$ tributaries |
| … | … |
| $(J-1)^{th}$ level | 2 tributaries |
| $J^{th}$ level | 1tributary(scale)+1tributary (approximation) |

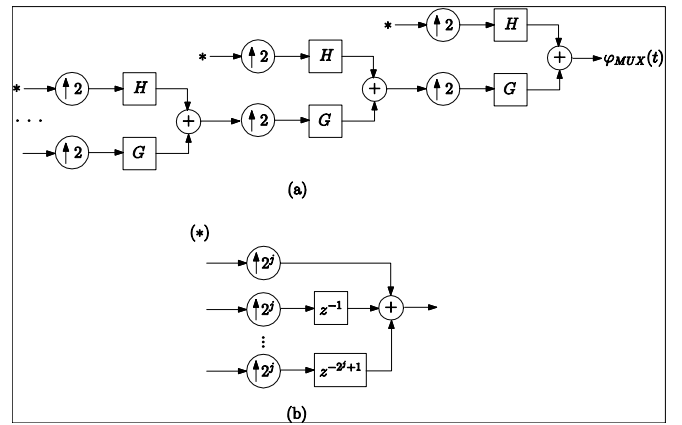

Fig. 1. (a) Three-level MRDM system. $G$ and $H$ are the scale and wavelet filters, respectively. (b) extra hardware required to implement TDM detail scales; in adjacent scales, the number of up-sampling blocks is scaled by a factor of 2, as well as the up-sampling factor, an so on.

The actual system make-up resembles a standard MRA: MRA synthesis performs multiplex, and MRA analysis performs demultiplex (Fig. 2).

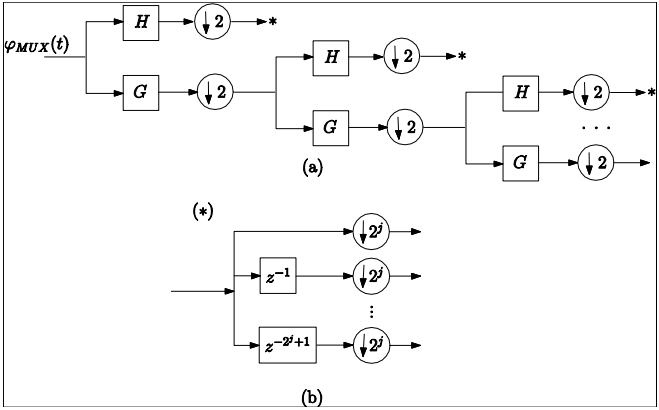

Fig. 2. (a) MRA-based demultiplex system carried out with filters *G* and *H*. (b) The extra hardware required to separate information from different users at a given detail channel.

## III. CHANNEL WAVEFORMS AND SPECTRAL SHAPE

The channel waveforms at the multiplex output are strongly depending upon the wavelet system selected in the multiplex process. A few spectra are presented for the Haar-based MRDM and two scenarios are offered. First, the number of scales of the pseudo-MRA is maintained constant, and the spectrum changes caused by increasing the MRA blocklength is evaluated (Fig.3).

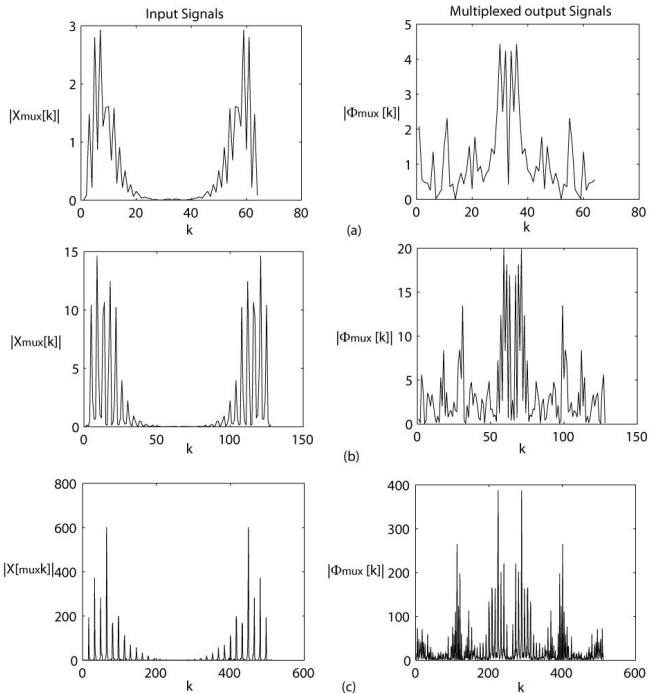

Fig. 3. Spectral effects of increasing the user rate in a Haar-MRDM with a fixed number of $n_J=2^J=4$ users. The number of decomposition levels is $J=2$. *N* is the MRA blocklength. (a) *N*=64, $N/2^J$=16 per user. (b) *N*=128, $N/2^J$=32 per user. (c) *N*=512, $N/2^J$=128 per user.

In all cases, data from tributaries are assumed to have the same bit rate. A signal $x_{mux}(t)$ was then generate, which corresponds to the concatenation of user signals. Its discrete time version $x_{mux}[n]$ is a TDMed version that has the same length as the multiplexed signal $\varphi_{mux}[n]$. Plots show the magnitude of spectra of the time division multiplexed signal $|X_{mux}[k]|$ and the MRA signal $|\Phi_{mux}[k]|$, $k=1,2,…,N$. The effects of increasing the number of decomposition levels *J* for a Haar-based MRDM is also investigated (Fig.4).

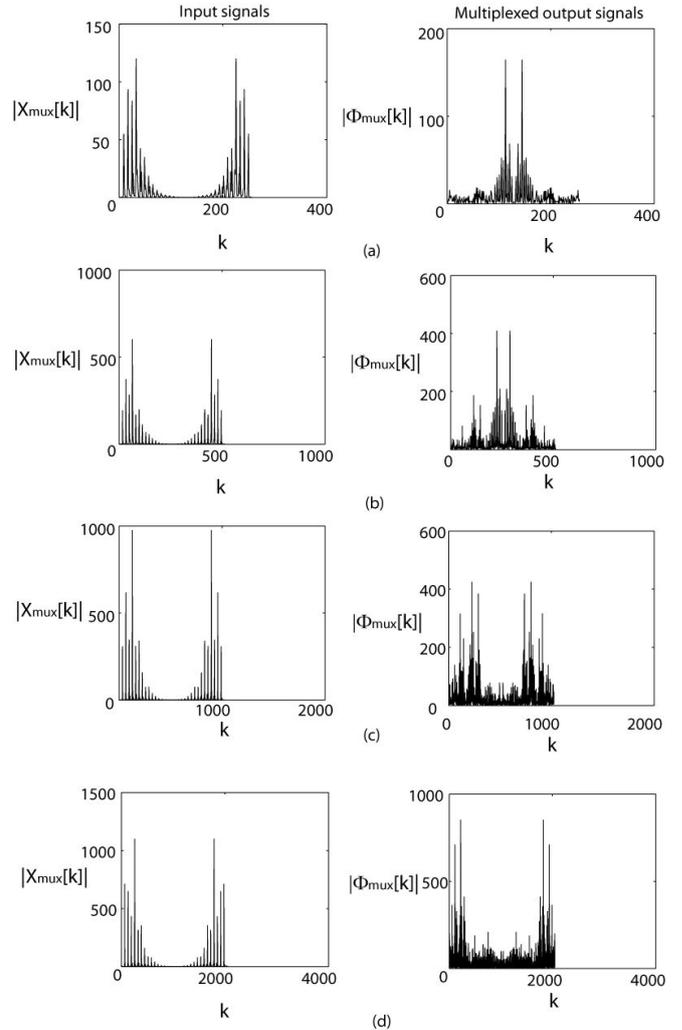

Fig. 4. Spectral effects of increasing the number of decomposition multiresolution levels rate in a Haar-MRDM. (a) *J*=2 ($n_J$=4 channels), *N*=256, $N/2^J$=64 (b) *J*=3 ($n_J$=8 channels), *N*=512, $N/2^J$=64. (c) *J*=4 ($n_J$=16 channels), *N*=1024, $N/2^J$=64. (d) *J*=5 ($n_J$=32 channels), *N*=1048, $N/2^J$=64.

## IV. MULTIRATE MRDM SYSTEMS DISSECTED

This section presents a few naïve cases of multiplex in order to shed light on the operation of MR-based multiplexes. In a more general approach, different users can require different rates. At a given scale, different users can be primary multiplexed via a standard synchronous time division multiplex (TDM) so as to engender the signal transmitted at this level. Let *J* denotes the number of scales in the pseudo-MRA, and let *N* be the blocklength of the MRA. The multiplex assume a bit rate *R* for the basic channel (a reference rate). The main parameters of the MRDM are therefore (*N*, *J*, *R*).

For analog input signals, the multiplex can handle with signals of maximum frequency $f_m$. Each signal digitalised using a *B*-bit A/D converter, so the transmission bit rate is $R=B \cdot 2f_m$ bps. The value of the wavelet coefficients $f_j[k]$ is set according to a sample of signal of the $j^{th}$-channel. The signal at the coarsest decomposition level corresponds to that of lower bit rate. Therefore, *T* (the time required for the transmission of the multiplexed frame) is given by

$$T = \frac{N \cdot B}{2^J \cdot R} \text{ (s)}. \qquad (4)$$

Tributary channels at different bit rates can be multiplexed in these MRDM schemes. Tributary rates expressed in bps range from $R_{min}$ to $R_{max}$, where:

$$R_{min}=R, 2R, 4R, \ldots, 2^{J-1}R=R_{max}. \quad (5)$$

Let $n_j$ denote the number of channels at a rate $2^{J-j}R$. Then

$$\sum_{j=1}^{J} \frac{n_j}{2^j} = 1. \quad (6)$$

Consequently, the maximum number of users is $n_J=2^J$ at the same rate $R$ (bps). This is precisely the case examined in the section II and III (all users with equal rate).

A naïve example presents a basic channel limited at $f_m=4$ kHz (e.g. a voice channel) using an 8-bit PCM ($R=64$ kbps – in the case of digital input, i.e. a B-channel). Assuming an MRDM with $J=3$ and $N=512$, the time of a frame is $T=8$ ms. This is denoted by MRDM (512, 3, 64 kbps). A more efficient scheme able to multiplex the same number of channels can be implemented with a MRDM (64,3,64 kbps) using a frame of $T=1$ ms and 8 bit/sample A/D converter.

TABLE I
POSSIBLE COMPOSITION IN AN MRDM (64,3,64 kbps): THE ASSIGNMENT $A_I$ SHOWS THE NUMBER OF MULTIPLEXED CHANNELS OF EACH KIND.

| multiplex make-up | 256 kbps $n_1=$ | 128 kbps $n_2=$ | 64 kbps $n_3=$ |
|---|---|---|---|
| $A_1$ | 2 | 0 | 0 |
| $A_2$ | 1 | 2 | 0 |
| $A_3$ | 1 | 1 | 2 |
| $A_4$ | 1 | 0 | 4 |
| $A_5$ | 0 | 4 | 0 |
| $A_6$ | 0 | 3 | 2 |
| $A_7$ | 0 | 2 | 4 |
| $A_8$ | 0 | 1 | 6 |
| $A_9$ | 0 | 0 | 8 |

Therefore, the MRDM can deal with one of the following configuration (Table I): multiplex two users at 256 kbps; or multiplex a single 256 kbps user and a pair of 128 kbps users; … multiplex channel at different rate (64, 128 and 256 kbps); …; or even multiplex eight identical 64 kpbs synchronous tributaries. Table I values can be as well used for the MRDM (512, 3, 64 kbps). The sole difference is the frame duration. Figures 5, 6, 7 and 8 illustrate a number of the up-and-coming configurations. They correspond to the set-up $A_1, A_2, \ldots, A_4$ described in Table I, respectively.

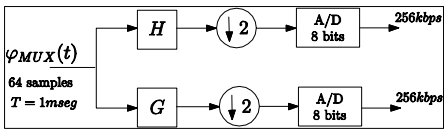

Fig. 5. $A_1$-Type multiplex: in this arrangement, two users are retrieved with transmission rates equal to 256 kbps.

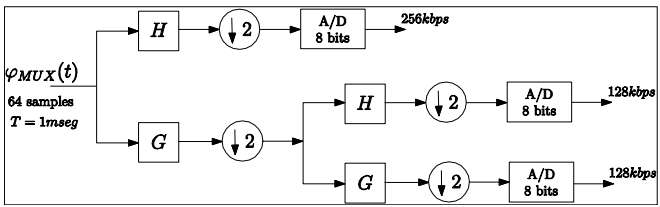

Fig. 6. $A_2$-Type multiplex: three users are retrieved, namely two users at 128 kbps and one user with transmission rate 256 kbps.

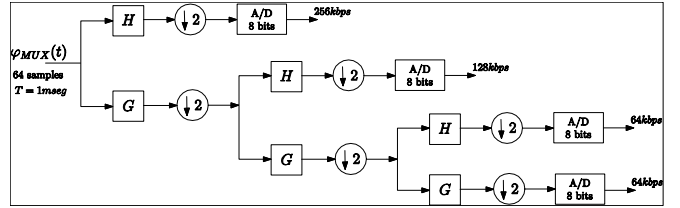

Fig. 7. $A_3$-Type multiplex: four users recovered with transmission rates equal to 256 kbps, 128 kbps and 64 kbps, respectively.

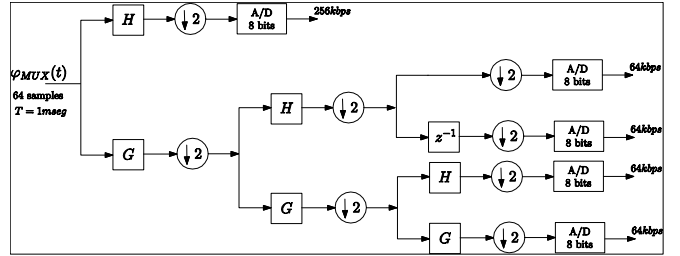

Fig. 8. $A_4$-Type multiplex: five users are demultiplexed with transmission rates equal to 256 kbps and 64 kbps, respectively. Remark the presence of the extra hardware at the second decomposition level, required to demultiplex two previously TDMed users.

As another straightforward example, let us consider the MRDM (256, 5, 64 kbps), which uses a frame of $T=125$ μs and 1-bit D/A converters. The multiplex output rate is 2.048 Mbps so it can be used to multiplex channels in the integrated services digital network ISDN at primary rate user-network (G.704). Further promising configurations are MRDM (128, 5, 64 kbps) with 2-bits converters, MRDM (64, 5, 64 kbps) with 4-bits converters or MRDM (32, 5, 64 kbps) with 8-bits converters. In all cases, $T=125$ μs (PCM-30 time frame) and output rate is maintained at 2.048 Mbps (E1 frame). It should be kept on mind that these are merely minor examples of multiplexing, but sophisticated systems with a great number of users can as well as be easily implemented.

## V. CONCLUSIONS

This paper introduces the groundwork of novel SP-based schemes, which are offered as a choice for multiplexing analog or digital signals. In this class of multiplex, the users are separated in a "scale domain". Both equal and multirate users are considered. As expected, bandwidth requirements are tantamount to those of FDM, TDM or CDM systems. The major advantage of this technique is that both multiplex and demultiplex can efficiently be assisted by signal processors (DSP), and no specialised hardware is required to implement it. Different orthogonal wavelet systems can accomplish MRDM ensuing different channel waveform, and as a result several ways to fulfil the channel spectrum.

Further details on the functioning of this scheme, including synchronisation issues and performance evaluation over noisy channels are currently under investigation.